\documentstyle[11pt,newpasp,twoside,epsf]{article} 
\def\edcomment#1{\iffalse\marginpar{\raggedright\sl#1\/}\else\relax\fi} 
\reversemarginpar

\begin{document} 

\title{Constraints on the High Energy Continuum of Seyfert galaxies
from RXTE Color-Color and Color-Flux Diagrams}

\author{P.O. Petrucci} \affil{Osservatorio Astronomico di Brera,
Milano, Italy} \author{I. Papadakis$^{2}$, F. Haardt$^{3}$,
L. Maraschi$^{1}$, I. McHardy$^{4}$, P. Uttley$^{4}$}

\affil{$^1$Osservatorio Astronomico di Brera, Milano, Italy,
$^2$Physics Department, University of Crete, Heraklion, Greece;
$^3$Physics Department, Universita dell'Insubria, Como, Italy;
$^4$Department of Physics and Astronomy, University of Southampton, UK}

\begin{abstract} 
We report on some results from a three-year program of RXTE
observations of 4 Seyfert galaxies: NGC~4051, NGC~5548, MCG-6-30-15
and NGC~5506. We focus here on color-color and color-flux diagrams
obtained using the count-rates of the PCA instrument in different
energy ranges: 3--5, 5--7 and 7--10 keV.\\ The data of the different
sources show interesting trends in the different color-color and
color-flux planes. These trends are quite similar from one source to
the other, even in the case of the narrow line Seyfert 1 NGC~4051
which exhibits strong flux and spectral variability.\\ We discuss
these results in term of a simple spectral model: a power law with
high energy cut-off + reflection + Iron line, with the scope of
understanding the relations between the observed variations of
different components and the physical causes of the variability.
\end{abstract} 

The main aim of the project was to study the X--ray flux variability
of NGC~4051, NGC~5548, MCG-6-30-15 and NGC~5506 the sources on long
time scales (Uttley et al., in preparation). For that reason, we tried
to sample the largest possible range of the variability time scales
from sub--daily to yearly, so the objects were observed with a
different rate at different parts of the whole 3 years
period. Typically, each observation had an average exposure time of
$\sim$ 1 ksec. Here we focus on the spectral variability observed in
these 4 objects.
\section{The Data} 
\begin{table}
\begin{tabular}{lccc}
\tableline
Object Name& Start day & End day & Observations Nb\\
\tableline
NGC 5506        &07/05/96               &02/02/99& 127\\
NGC 5548        &23/04/96               &22/12/98& 134\\
MCG-6-30-15     &08/05/96               &02/02/99& 124\\
NGC 4051        &23/04/96               &02/02/96& 140\\
\tableline
\tableline
\end{tabular}
\caption{The RXTE observation log}
\end{table}
The data was reduced using FTOOLS v.4.2. The light curves include only
the PCA, top Xenon layer data, where the PCA is most sensitive. PCA
"good times" have been selected from standard 2 data using the
"normal" criteria for faint sources: "elv $>$ 10, time since SAA $>$
20 minutes, electron0,1,2 $<$ 0.1, offset $<$ 0.02 and PCUs 0,1 and 2
on" (the other PCUs were on and off occasionally and that is the
reason they were not included in the analysis).\\ All the light curves
include Epoch3 data ONLY, and the background estimation was done using
the latest L7 model for faint sources. We have reported in Table 1 the
observation period as well as the number of pointings for each
object.\\ In Fig. 1, we have plotted the 7-10 keV light curves of the
4 Seyfert galaxies. The Narrow-Line Seyfert galaxies NGC~4051 exhibits
larger amplitude than the other, as typical of this class of objects.
\begin{figure}[h]
\plotone{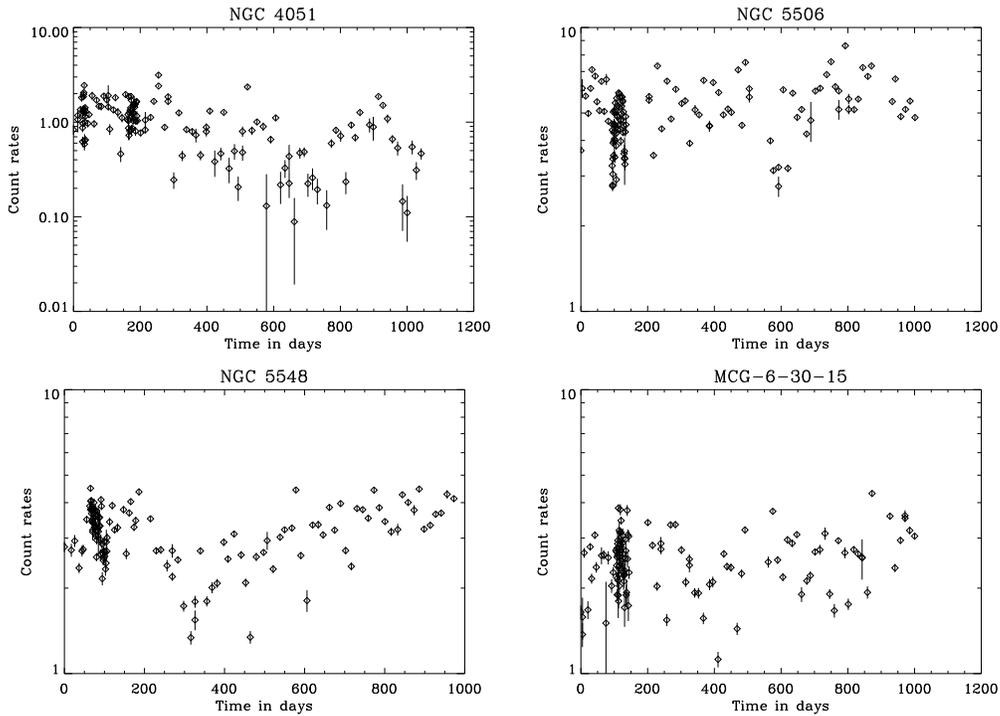}
\caption{7--10 keV light curves of the four objects of our sample
during the 3 year period of RXTE observations. Note the different
scales.}
\end{figure}

\section{The Method} 
\subsection{The Hardness Ratios} 
We use the count rates in four bands of the PCA instruments of RXTE:
3--5, 5--7 and 7--10 keV. We have then computed two different
hardness ratios HR1 and HR2 defined as follows:
\begin{equation}
\mbox{HR1}=\frac{(7-10\ keV)}{(3-5\ keV)},\ \mbox{HR2}=\frac{(7-10\
keV)}{(5-7\ keV)}
\end{equation}
HR1 is thus sensitive to the variability of the continuum and HR2 to
the variability of the equivalent width (EW) of the neutral Iron line
expected near 6.4 keV.

\subsection{The Model} 
We have computed the color-color and color-flux relations expected
with a cut-off power law + reflection + neutral Iron line model. We
have used the PEXRAV model (Magdziarz \& Zdziarski, 1995) of XSPEC
v10.0.\\ We have computed the hardness ratios with different values of
the spectral index ($\alpha$ between 0.3 and 1.6), reflection
normalization (R=0, 1, 2) and Iron line EW (0, 250 and 500 eV). The
column density for NGC 5506 being large ($N_h\simeq 10^{22}\ cm^{-2}$,
it was also included in the computation for this object.

\section{Results and Discussion} 
\subsection{The Color-Flux Diagrams: variability of the continuum shape} 
\begin{figure}[h]
\plotone{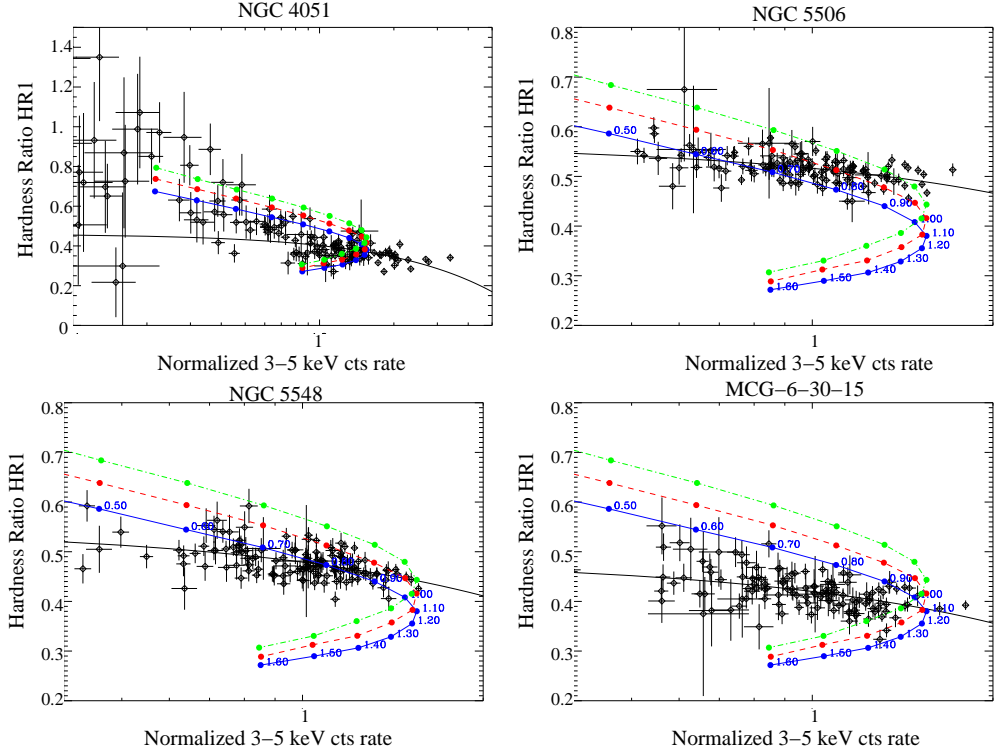}
\caption{HR1 vs (3-5) keV color-flux plots for each objects of our
sample. The black solid lines are the best linear fits (parameters
reported in Table 2). We have also over-plotted the color-flux trends
predicted by PEXRAV for different reflection normalization (from top
to bottom R=0(blue line), R=1 (red line), R=2 (green line)), {\bf
assuming a constant broad band (1-1000 keV) flux}. The photon index is
reported on the blue curves.}
\end{figure}
The color-flux plots HR1 versus the (3--5 keV) count rates of the four
galaxies are shown in Fig. 2. They have quite similar trends even in
the case of the Narrow-Line Seyfert 1 NGC~4051: {\bf the spectrum
softens when the flux increases}. The objects show spectral variation
$\Delta\alpha\simeq 0.3$ for flux variations of factor of two. At very
low flux however, NGC~4051 exhibits a different behavior.\\ We obtain
{\bf significant correlations} using the Spearman rank-order test. The
corresponding Spearman coefficients $r_s$ are reported, for each
object, in Table 2, with the best fit linear parameter (the solid
black lines in Fig. 2). Roughly similar slopes are found for each
object.\\ For comparison, we have over-plotted in Fig. 2 the hardness
ratios computed with the PEXRAV model, assuming a variation of the
spectral index with constant broad band (1-1000 keV) flux. Different
curves correspond to different values of the reflection normalization
R (0, 1 or 2).\\ We see that the color-flux trends are relatively well
explained by variation of the photon index at a constant broad band
flux.
\begin{table}
\begin{tabular}{lccccc}
\tableline
Object Name&$r_s$&$P$&a&b&$\chi^2_{\nu}$\\
\tableline
NGC 5506&--0.52&$\sim 10^{-10}$&-0.050$\pm$0.005&0.57$\pm$0.006&278/127\\
NGC 5548&--0.51&$\sim 10^{-10}$&-0.068$\pm$0.007&0.547$\pm$0.008&259/134\\
MCG-6-30-15&--0.49&$\sim 10^{-9}$&-0.064$\pm$0.007&0.484$\pm$0.008&244/124\\
NGC 4051&--0.70&$\sim 10^{-19}$&-0.057$\pm$0.005&0.460$\pm$0.008&310/118\\
\tableline
\tableline
\end{tabular}
\caption{Linear best fits (i.e. y=ax+b) of the color-flux diagram HR1
vs. (3--5) keV plotted in Fig. 2. Errors in both coordinates are taken
into account. We also list in the first two columns the Spearman
rank-order correlation coefficient $r_s$ and its significance $P$.}
\end{table}

\section{The Color-Color Diagrams: Variability of the Iron Line} 
\begin{figure}[h]
\plotone{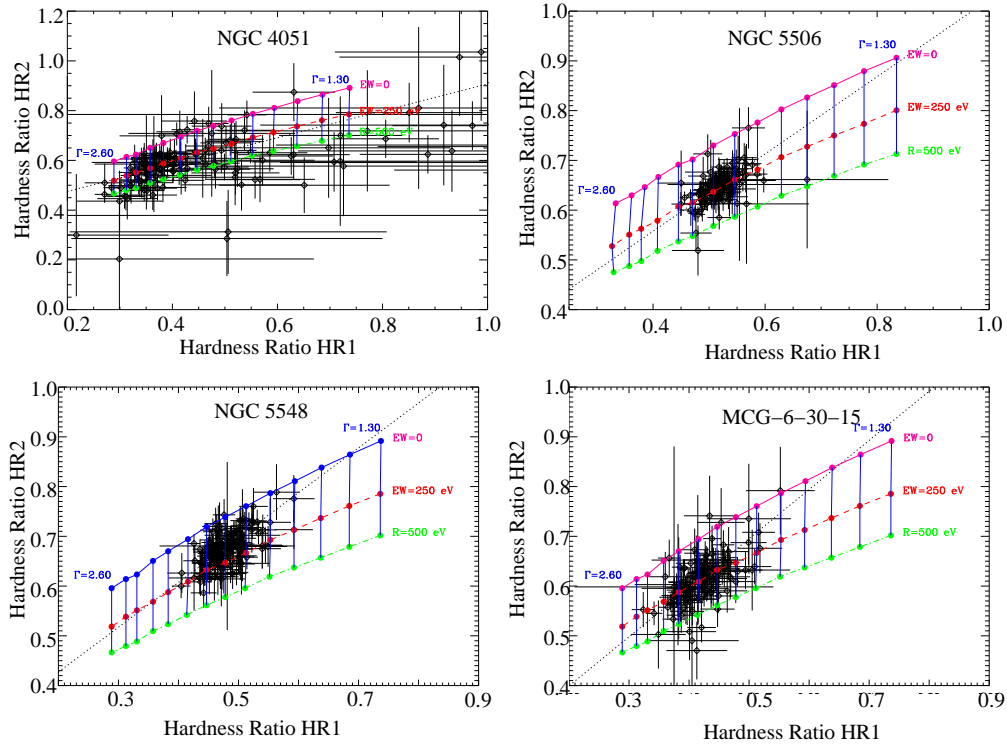}
\caption{HR2 vs HR1 color-color plots for each object. The black
dotted lines are the best linear fits (parameters reported in Table
3). We have also over--plotted the hardness ratios predicted by PEXRAV
for different photon index (blue solid vertical lines) and line EW (0
eV (pink line), 250 eV (red line), 500 eV (green line)).}
\end{figure}
\begin{table}[h!]
\begin{tabular}{llccc}
\tableline
Color-color&Object Name&a&b&$\chi^2_{\nu}$\\
\tableline
HR2-HR1&NGC 5506&0.78$\pm$0.09&0.25$\pm$0.05&92/127\\
&NGC 5548&0.90$\pm$0.11&0.25$\pm$0.05&93/134\\
&MCG-6-30-15&0.99$\pm$0.12&0.20$\pm$0.05&98/124\\
&NGC 4051&0.54$\pm$0.08&0.37$\pm$0.03&132/140\\
\tableline
\tableline
\end{tabular}
\caption{ Linear best fits (i.e. y=ax+b) of the different color-color
diagrams plotted in Fig. 3. Errors in both coordinates are taken into
account.}
\end{table}
We have plotted, in Fig. 3, the color--color diagrams HR2-HR1 for each
object of our sample. We have also reported the hardness ratios
predicted by PEXRAV for different photon index and line EW. It appears
that:
\begin{itemize}
\item The two hardness ratios are roughly proportional (best linear
fit parameters reported in Table 3) meaning that the line flux follows
the continuum variability.
\item The line EW keeps quite constant during spectral changes. For
harder spectra, the EW slightly decreases.
\end{itemize}

\section{Conclusion} 
\begin{itemize}
\item All four objects of our sample exhibit flux and spectral
variabilities, the larger ones being observed in the Narrow Line
Seyfert galaxies NGC~4051, as typical of this class of objects..
\item The color-flux diagrams show systematic trends, quite similar
for each object, even in the case of NGC~4051 (at large flux): {\bf
the spectra always soften when the flux increases}.
\item The color-flux trends are relatively well explained by {\bf
variation of the spectral index (between $\sim$0.6 and $\sim$1.1) at
constant broad band flux.}
\item The {\bf EW of the Iron line keeps roughly constant} during
spectral variability meaning that the Iron line flux follows the
continuum. This result supports the idea that the Iron line is
produced by reflection from the matter which is very close to the
X--ray emitting region.
\end{itemize}


\end{document}